\documentclass[a4paper,11pt]{article}
\usepackage{pos}
\usepackage{bm}
\usepackage{multicol}

\title{Spectral reconstruction in NRQCD via the Backus-Gilbert method}
\author*[a]{Ben Page}
\author[a,b]{Gert Aarts}
\author[a]{Chris Allton}
\author[c]{Benjamin J\"ager}
\author[d]{Seyong Kim}
\author[e]{Maria Paola Lombardo}
\author[a]{Sam Offler}
\author[f]{Sinead M. Ryan}
\author[g]{Jon-Ivar Skullerud}
\author[a]{Thomas Spriggs}

\affiliation[a]{Department of Physics, Swansea University, Swansea SA2 8PP, United Kingdom}

\affiliation[b]{European Centre for Theoretical Studies in Nuclear Physics and Related Areas (ECT*) \& Fondazione Bruno Kessler Strada delle Tabarelle 286, 38123 Villazzano (TN), Italy}

\affiliation[c]{CP3-Origins \& Danish IAS, Department of Mathematics and Computer Science, University of
Southern Denmark, 5230 Odense M, Denmark}

\affiliation[d]{Department of Physics, Sejong University, Seoul 143-747, Korea}

\affiliation[e]{INFN, Sezione di Firenze, 50019 Sesto Fiorentino (FI), Italy}

\affiliation[f]{School of Mathematics, Trinity College, Dublin 2, Ireland}

\affiliation[g]{Dept. of Theoretical Physics, National University of Ireland Maynooth, County Kildare, Ireland}

\emailAdd{b.page.900727@swansea.ac.uk}

\abstract{
We present progress results from the {\sc Fastsum} collaboration's 
programme to determine the spectrum of the bottomonium 
system as a function of temperature using a variety of
approaches.
In this contribution, the Backus-Gilbert method is used to reconstruct spectral functions from NRQCD meson correlator data from {\sc Fastsum}'s anisotropic ensembles at nonzero temperature. We focus in particular on the resolving power of the method, providing a demonstration of how the underlying resolution functions can be probed by exploiting the Laplacian nature of the NRQCD kernel. We conclude with estimates of the bottomonium ground state mass and widths at nonzero temperature.
}

\FullConference{%
 The 38th International Symposium on Lattice Field Theory, LATTICE2021
  26th-30th July, 2021
  Zoom/Gather@Massachusetts Institute of Technology
}

\begin{document}
\maketitle

\section{Introduction}

Bottomonium states play a special role in QCD at high temperature. They are produced copiously in current relativistic heavy-ion collision experiments
and can act as probes of the quark-gluon plasma, as their masses are much larger than other energy scales, including the temperature.

Spectral studies of bottomonium systems at nonzero temperature using lattice QCD are challenging due to the ``ill-posed'' nature of spectral reconstruction.
The {\sc Fastsum} Collaboration has studied bottomonium using lattice QCD at nonzero temperature on anisotropic lattices for some time \cite{Aarts:2010ek,Aarts:2011sm,Aarts:2014cda}.
Here we discuss attempts to study the bottomonium spectrum using our thermal, anisotropic lattices using a variety of different methods.
In this contribution the Backus-Gilbert method is used;
in other contributions the Maximum Likelihood \cite{Tom-Lat21} 
and Kernel Ridge Regression \cite{Sam-Lat21} are applied.

\section{Lattice details}

The large mass of the $b$ quark means that $M_b\gg T$ and so the 
nonrelativistic QCD (NRQCD) effective field theory can be used to 
study bottomonium mesons.
In this approximation, the Lagrangian is expanded in powers of the 
$b$ quark velocity, where expansions up to $\mathcal{O}(v^4)$ are 
sufficient to describe the behaviour of the $b$ \cite{Lepage1992}. 
Because the $b$-quark and antiquark decouple, the time evolution
of the $b$-quark propagator becomes an initial-value problem.

In NRQCD, the spectral representation of the Euclidean meson
correlator, $G(\tau)$, is given by
\begin{equation}
    G(\tau;T)=\int_{\omega_\text{min}}^{\omega_\text{max}}
    \frac{d\omega}{2\pi}
    K(\tau, \omega)\rho(\omega;T)
    \label{eq:spectral_repr}
\end{equation}
where $\rho(\omega;T)$ is the spectral density at temperature $T=(a_\tau N_\tau)^{-1}$ as a function of the energy $\omega$ and $K(\tau, \omega)$ is the temperature independent kernel of NRQCD, $K(\tau, \omega)=e^{-\omega\tau}$.
We note that there is an additive energy rescaling inherent in NRQCD,
and so the $\omega$ energy range
is related to the physical energy via
\begin{equation}
    E^\text{physical} = \omega + \Delta E, \qquad\qquad \Delta E= 7.46\, \text{GeV}.
    \label{eq:ephys}
\end{equation}
This analysis makes use of {\sc Fastsum}'s Generation 2L ensembles: anisotropic lattices ($\xi = a_s/a_\tau \sim 3.5$) with 2+1 flavour, clover-improved Wilson fermions using a  physical $s$ quark and lighter, degenerate $u$ and $d$ quarks
(see \cite{Aarts2020} for details).
The spatial extent of the lattice $N_s=32$ and there are $\mathcal{O}(1000)$ configurations at each temperature.
Details of the temperatures studied and the corresponding temporal extent, $N_\tau$, are listed in Table \ref{tab:Ntau_correlators}.

\begin{table}[h]
    \centering
 \begin{tabular}{|c || c | c | c | c | c | c | c | c | c | c | c |} 
 \hline
  $N_{\tau}$ & 128 & 64 & 56 & 48 & 40 & 36 & 32 & 28 & 24 & 20 & 16\\ [0.5ex] 
 \hline
  $T=1/(a_\tau N_\tau)$ [MeV] & 47 & 94 & 107 & 125 & 150 & 167 & 187  & 214 & 250 & 300 & 375\\
 \hline 
\end{tabular}
 \caption{Temporal extent and temperature in MeV for the {\sc Fastsum} Generation 2L ensembles \cite{Aarts2020}.}
    \label{tab:Ntau_correlators}
\end{table}

\section{The Backus-Gilbert method
\label{sec:bg}}
To solve Eq.~\eqref{eq:spectral_repr} for $\rho(\omega)$, we first
note that $G(\tau)$ is usually known at only
$\mathcal{O}(10)-\mathcal{O}(100)$ points, whereas it would
require $\mathcal{O}(1000)$ points to correctly represent
the continuous function $\rho(\omega)$.
This illustrates the ``ill-posed'' nature of this inverse problem.

Backus and Gilbert introduced their method for solving this problem
in the context of inverting gross Earth data in 1968 \cite{Backus1968}.
Their method's estimate, $\hat{\rho}$, of the solution to Eq.~\eqref{eq:spectral_repr} is generated from the target spectrum $\rho$ via a set of resolution functions $A(\omega,\omega_0)$,
\begin{equation}
    \hat{\rho}(\omega_0) = \int_{\omega_\text{min}}^{\omega_\text{max}} A(\omega,\omega_0)\rho(\omega)d\omega.
        \label{eq:bg}
\end{equation}
Ideally $A(\omega,\omega_0)$ closely approximates the delta function
$\delta(\omega-\omega_0)$.
The key point is that the resolution functions are a linear
combination of the kernel,
\begin{equation}
    A(\omega,\omega_0) = \sum_\tau c_\tau(\omega_0) K(\tau, \omega).
    \label{eq:res-fns}
\end{equation}
where $c_\tau(\omega_0)$ are the yet-to-be-determined Backus-Gilbert coefficients.
Combining Eqs.~\eqref{eq:bg} and \eqref{eq:res-fns}, the spectrum estimate $\hat{\rho}$ can now be expressed
linearly in terms of the correlation function,
\begin{equation}
    \hat{\rho}(\omega_0) = \sum_\tau c_\tau(\omega_0) G(\tau).
\end{equation}
We use the ``least-squares'' or ``Dirichlet'' approach
to determine the coefficients $c_\tau(\omega_0)$. This
minimises the distance $J$
between the resolution functions and the delta function
\cite{Oldenburg1984},
\begin{equation}
    J(\omega_0) = \int_{\omega_\text{min}}^{\omega_\text{max}} \left[A(\omega,\omega_0)-\delta(\omega-\omega_0)\right]^2~d\omega.
    \label{eq:dirichlet criterion}
\end{equation}
The coefficients $c_\tau (\omega_0)$ corresponding to the minimisation of Eq.~\eqref{eq:dirichlet criterion} can be found by solving the matrix-vector product,
\begin{equation}
    \mathcal{K}_{\tau\tau'} \cdot c_{\tau'}(\omega_0)= K(\omega_0,\tau),\qquad \text{where}\quad \mathcal{K}_{\tau\tau'}= \int_{\omega_\text{min}}^{\omega_\text{max}} K(\tau, \omega) K(\tau', \omega) d\omega.
    \label{eq:dirichlet matrix}
\end{equation}
%
%
%
The kernel width matrix $\mathcal{K}$ is almost singular and
it is therefore necessary to impose a regularisation routine in order to invert Eq.~\eqref{eq:dirichlet matrix}.
This is achieved by adding the covariance matrix $\Sigma$ of the underlying data,
$G(\tau)$, to $\mathcal{K}$  \cite{Brandt2015},
\begin{equation}
    \mathcal{K}(\alpha) = \alpha\mathcal{K} + (1-\alpha)\Sigma,
    \label{eq:white}
\end{equation}
where $\alpha\in(0,1]$ is the parameter which determines the strength of the regularisation.

A consequence of this redefinition is that there is a counter-play
between the resolution width,
i.e.\ the ability of the method to resolve fine spectral features,
and the stability of the solution
\cite{Yanovskaya2003,Conrath1977}; the greater the regularisation,
the worse the method's resolving power.


\section{Results for the $\Upsilon$ meson}
\label{sec:overview}

\begin{figure}[h]
    \centering
    \includegraphics[width=0.7\linewidth]{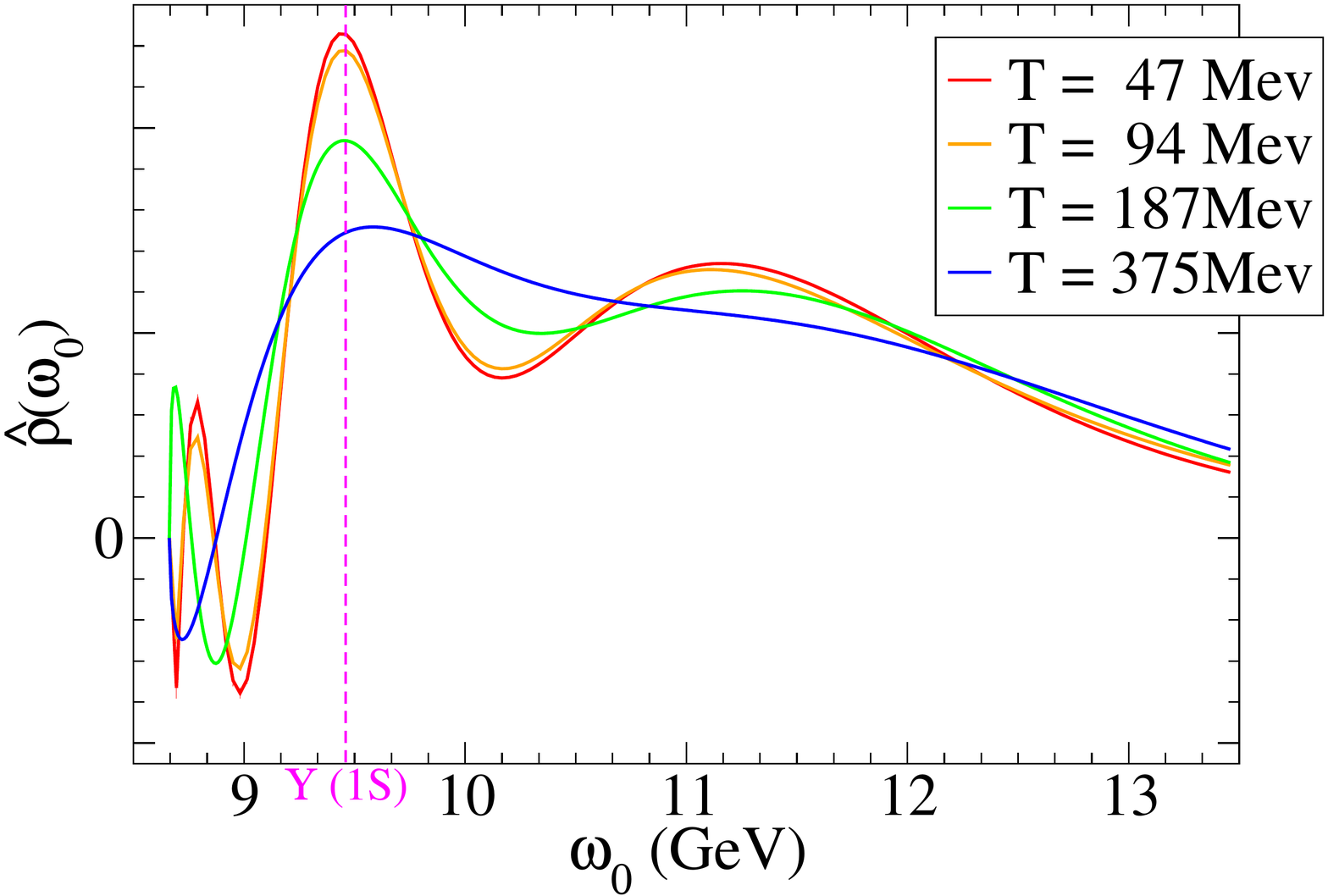}
    \caption{$\Upsilon$ spectrum as predicted by the  Backus-Gilbert method at four temperatures for the choice $\omega_\text{min}\approx8.66$ GeV and $\alpha=0.1$. Local quark sources were used. The magenta line is the experimental value of the $\Upsilon$ mass \cite{Zyla2020}. The statistical error in $\hat{\rho}(\omega_0)$, estimated via bootstrap resampling, is too small to be seen.}
    \label{fig:spp_i example}
\end{figure}

We use NRQCD correlation functions computed on {\sc Fastsum}'s anisotropic ensembles \cite{Aarts2020} to estimate the
$\Upsilon$ ground state mass, $M$, and width, $\sigma$
using the Backus-Gilbert method outlined above.
In Fig.\ref{fig:spp_i example} we present results for the Backus-Gilbert
$\Upsilon$ spectral function $\hat{\rho}$ at four 
temperatures.
In these plots, $\alpha=0.1$ (see Eq.\ (\ref{eq:white})) and $\omega_\text{min}=8.66$ GeV.
The vertical magenta dashed line shows the experimental value of the $\Upsilon$ mass.
As can be seen, the Backus-Gilbert estimate changes less as the 
temperature decreases, even though the number of $\tau$-points, and therefore the efficacy of the method increases as $T\rightarrow 0$.

The explicit dependence of $\hat\rho(\omega_0)$ on $\alpha$
and $\omega_\text{min}$ is unknown.
We therefore estimate this systematic error 
from an ensemble of $\mathcal{O}(500)$
choices of $\alpha$ and $\omega_\text{min}$,
with $\alpha$ ranging from 1.0 (no regularization) to $10^{-5}$
and $\omega_\text{min}$ ranging from $-0.2a_\tau^{-1}$ to $0.2a_\tau^{-1}$ (which in physical units is $\approx6.26$ GeV to $\approx8.66$ GeV, see Eq.~\eqref{eq:ephys}).
In addition, a bootstrap analysis using 1000 samples drawn from
the Monte Carlo ensemble of $G(\tau)$ is used to determine the
statistical error. The largest contribution to the error in $\hat{\rho}(\omega_0)$ is from systematic sources (see  Fig.~\ref{fig:laplace_wmin_comparison}). Fig.~\ref{fig:spp_i example} shows that the statistical error in $\hat{\rho}(\omega_0$) estimated by using samples $G(\tau)$ is small, although  these errors do become significant in the case of either Laplace shifting (see Fig.~\ref{fig:laplace shift}) or when there is little to no regularisation, see Eq.~(\ref{eq:white}).

The ground state feature is found by seeking the peak nearest
to the expected mass $\approx 9.4$ GeV.
The ground state mass $M$ and width $\sigma$ are then
extracted via a Gaussian fit to the top 50\% of the leading edge
(i.e.\ $\omega \le M$) of this peak.
The trailing edges (i.e.\ $\omega \ge M$) are not included in these
Gaussian fits because they tend to be contaminated by contributions
from higher energy states.
There are often ``side-lobe'' artefact features below this ground state. To ensure only robust estimates are included in the analysis, cases where these artefacts reach more than 50\% of the ground
state's peak are discarded.

The number of rejected fits increases as $T\rightarrow0$ due to 
small-$\omega$ oscillations in the construction of the 
$A(\omega,\omega_0)$.
For temperatures $T>50$ MeV ($N_\tau < 128$), the routine rejects 
$<10$\% of fits.
For the lowest temperature of $T\simeq 47$ MeV ($N_\tau = 128$), results 
from around 80\% of the tested parameter pairs were rejected. 

\begin{figure}[t]
    \centering
    \includegraphics[width=0.49\linewidth]{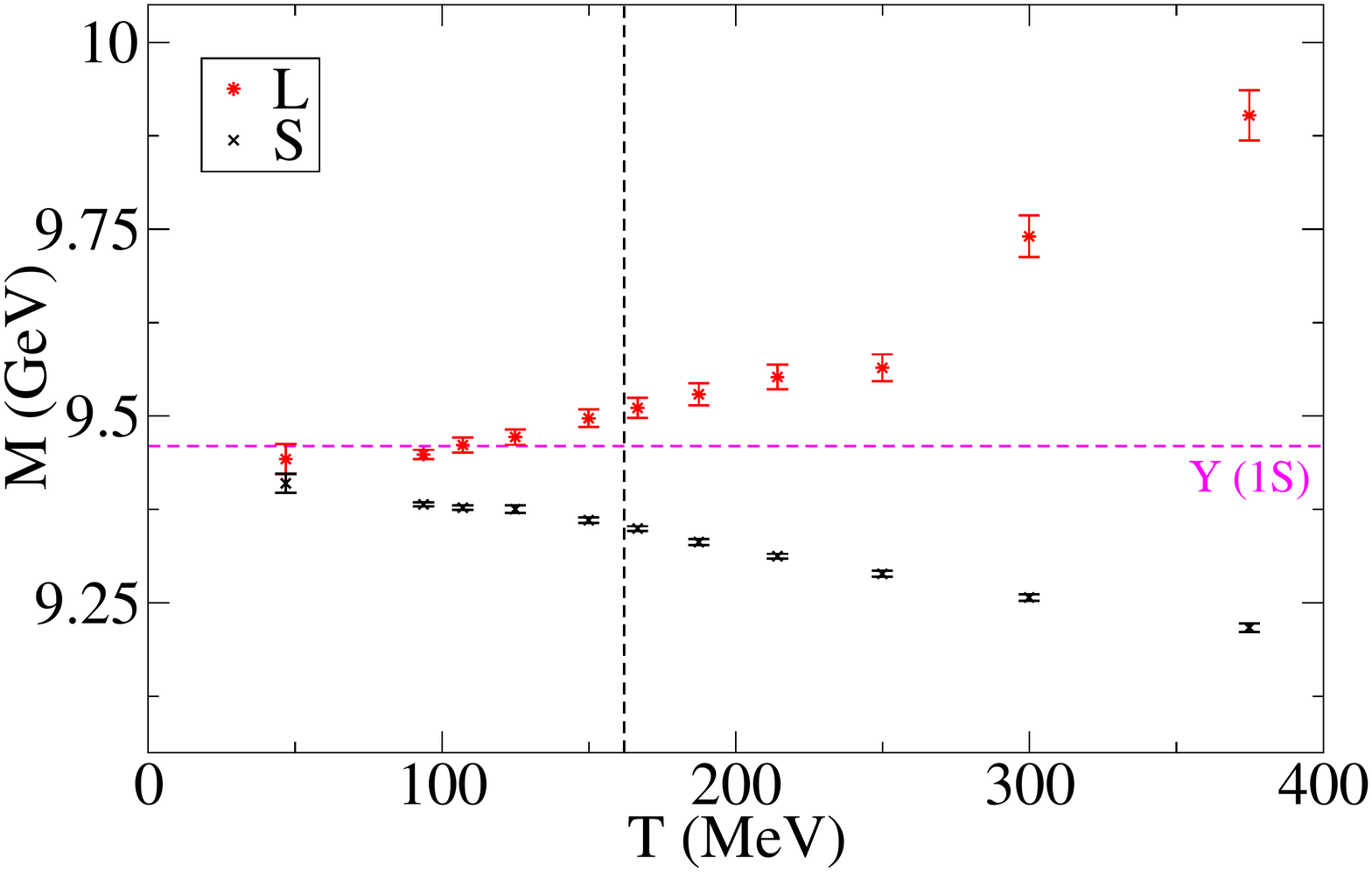}
    \includegraphics[width=0.49\linewidth]{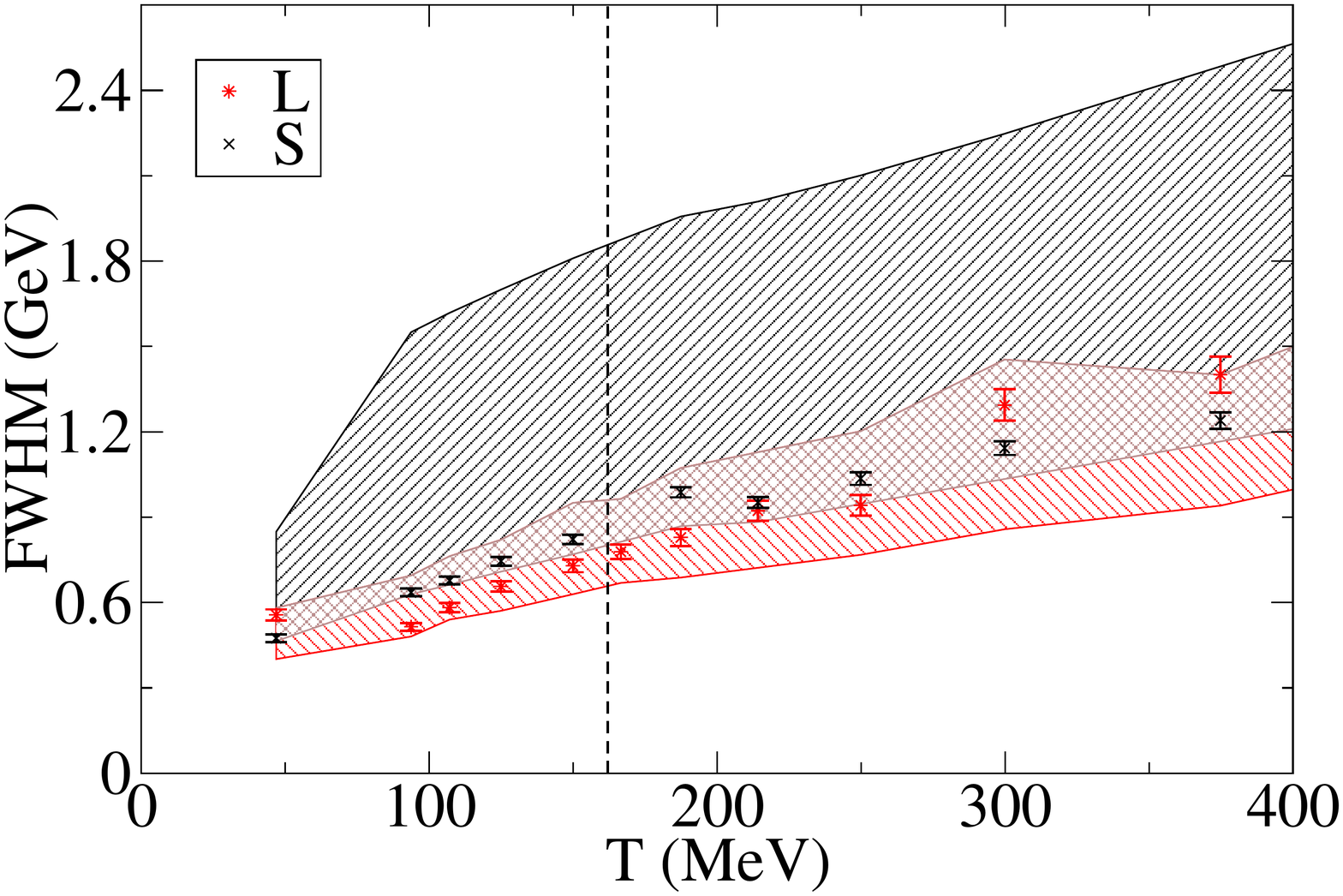}
    \caption{Ground state mass $M$ (left) and full width at half maximum (FWHM, right) versus temperature in the $\Upsilon$ channel with local (L) and smeared (S) sources. The horizontal magenta line
    is the experimental value for the $\Upsilon$ mass and the vertical black line is the pseudo-critical temperature, $T_{\rm pc}=162(1)$ MeV, determined by the inflection point of the renormalised chiral condensate \cite{Aarts2020}. The shaded regions indicate the minimum resolvable width for local (red) and smeared (black) sources, see below.
    }
    \label{fig:mass_width_comparison}
\end{figure}

Fig.~\ref{fig:mass_width_comparison} shows estimates for the ground state mass and width using this procedure as a function of temperature.
Both local and smeared quark sources are used where the smearing was chosen to have a good overlap with the ground state.
The ground state width appears independent of source type
for all $T$ considered. In the case of the ground state mass, the approach to the $T=0$ estimate differs between source types. Given this difference in behaviour, it is uncertain whether the ground state mass increases or decreases with increasing temperature.

\section{Systematic errors and improvements}
\label{sec:systematic_error}

{\bf Varying the Euclidean time window} -- 
At small $\tau$, $G(\tau)$ sees contributions from excited states.
One could imagine restricting the time window,
$[\tau_1,\tau_2]$, to include large Euclidean times only
where these excited states are exponentially suppressed.
However, restricting the $[\tau_1,\tau_2]$ interval
reduces the number of kernel functions used to construct the
resolution function $A(\omega,\omega_0)$ which consequentially
limits the resolving power of the method.
For this reason, our analysis includes the full time range,
$[\tau_1,\tau_2] = [0,N_\tau-1]$.

\begin{figure}[t]
    \centering
    \includegraphics[width=0.49\linewidth]{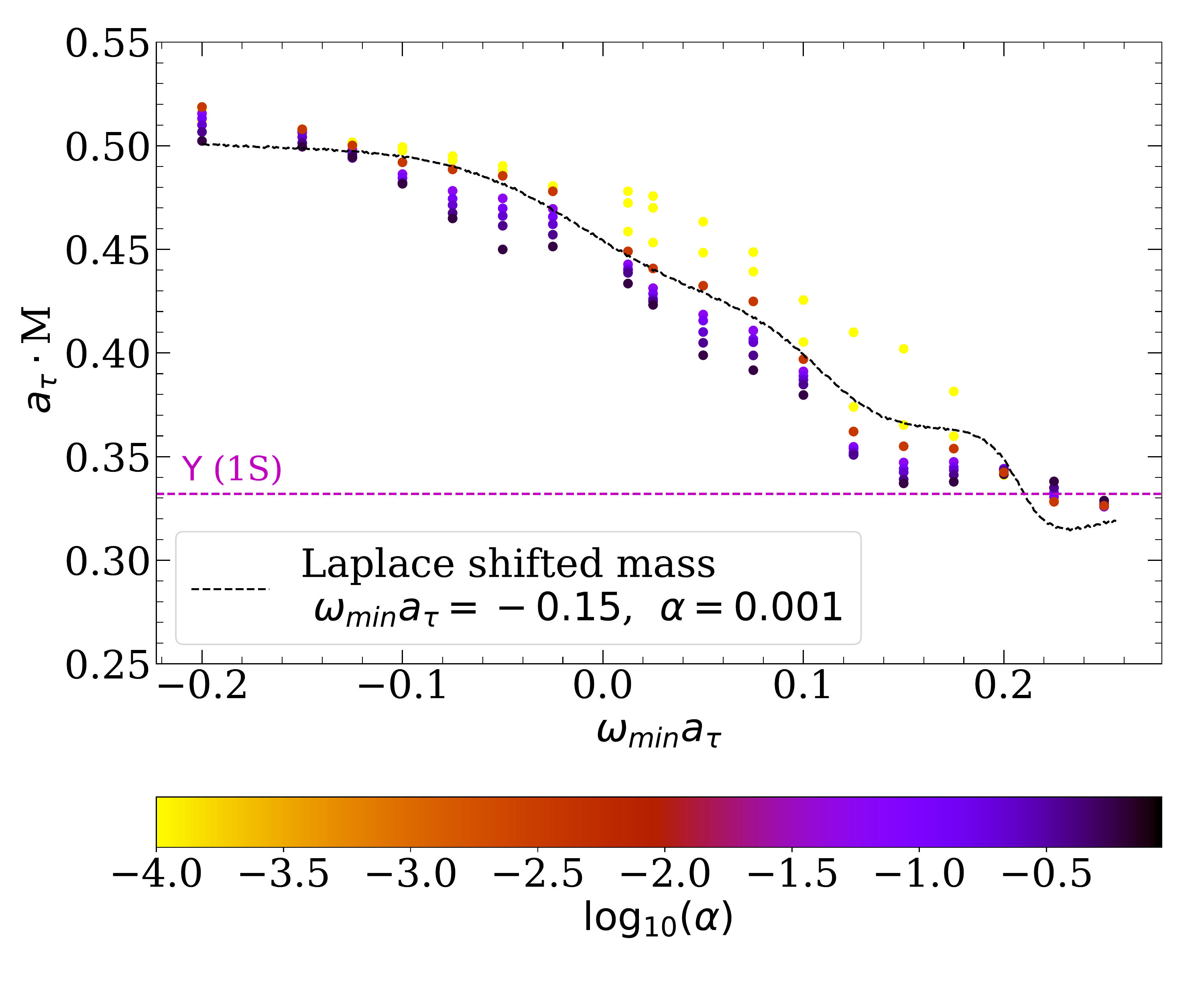}
    \includegraphics[width=0.49\linewidth]{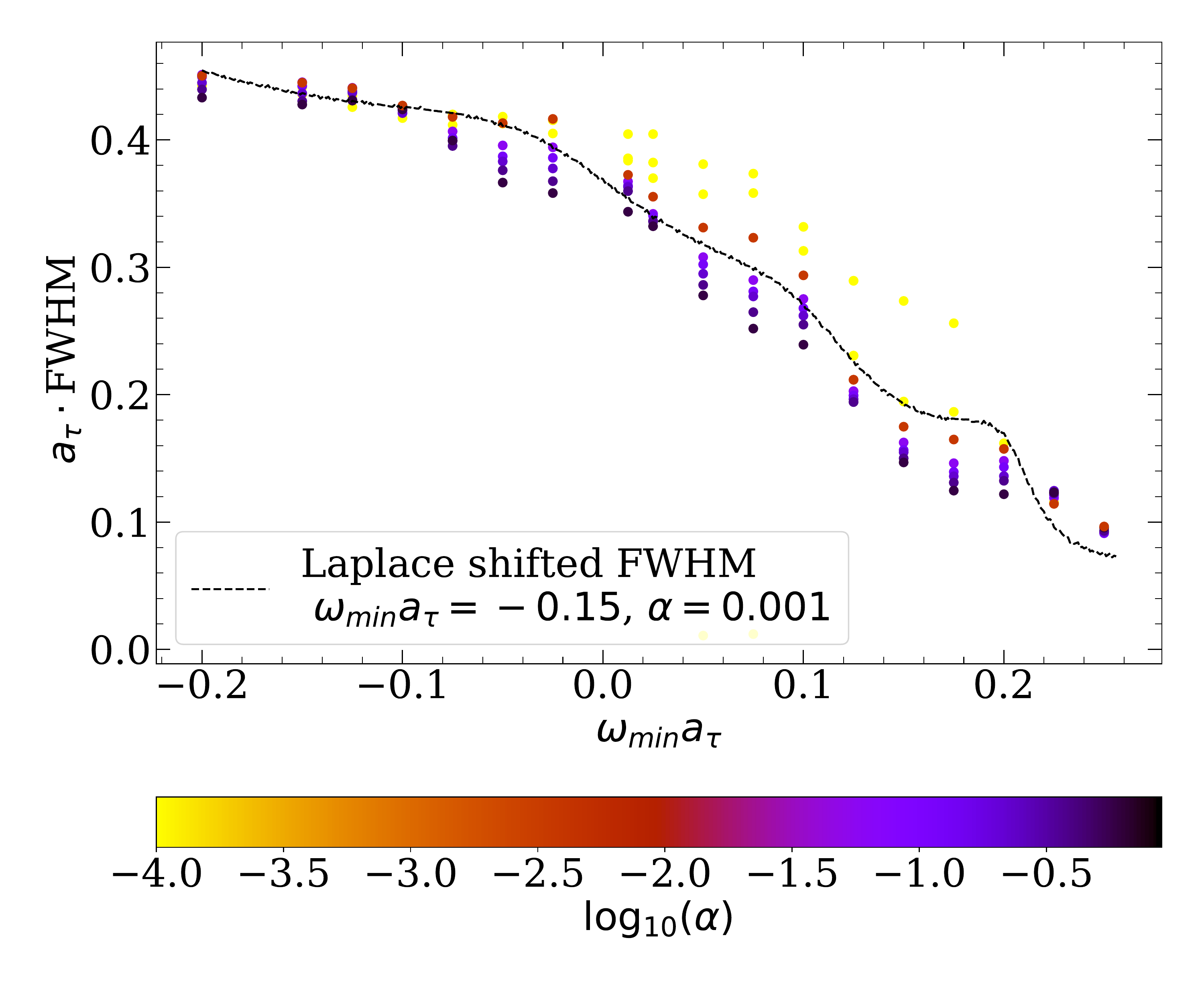}
    \caption{Scatter plot showing how the mass $M$ (left) and FWHM (right) of the ground state changes with $\omega_\text{min}$ and $\alpha$, indicating the systematic improvement in resolution. The horizontal magenta line is the experimental estimate of the $\Upsilon$ mass and $T=375$ MeV.
    In addition, in the case of Laplace shifts, the dashed line shows how $M$ and FWHM change with $\Delta$ for the case of $\omega_\text{min} a_\tau = -0.15$ and $\alpha=0.001$.  Note that the Laplace shifted $M$ and FWHM are plotted against $-0.15 + \Delta$ so that a direct comparison can be made with the $\omega_\text{min}$-varying case. This indicates a Laplace shift is equivalent to changing the lower bound $\omega_\text{min}$.
    }
    \label{fig:laplace_wmin_comparison}
\end{figure}

{\bf Varying $\omega_\text{min}$ and $\alpha$} --
As discussed in Sec.\ \ref{sec:overview}, 
we generate Backus-Gilbert estimates at multiple values of $\alpha$
and $\omega_\text{min}$ to probe the systematics related to
these two quantities.
We note that the resolving power of the Backus-Gilbert approach
depends on both $\omega_\text{min}$ and $\alpha$.
It is a feature of the Backus-Gilbert approach that its
resolution is best for features nearest $\omega_\text{min}$,
and therefore the resolution improves as $\omega_\text{min}$
increases.
Also, as discussed in Sec.\ \ref{sec:bg}, the resolution increases with
$\alpha$.
These two features are illustrated in Fig.\ \ref{fig:laplace_wmin_comparison}
where the mass and full width at half maximum (FWHM) are plotted against $\omega_\text{min}$.
Different values of $\alpha$ are shown using the colour coding indicated.
As can be seen the FWHM decreases (i.e.\ the resolution increases)
with $\omega_\text{min}$. The same is observed as $\alpha$ increases,
for fixed $\omega_\text{min}$.


{\bf Laplace shifting and noise subtraction} -- 
The exponential nature of the NRQCD kernel means that the spectral representation of the Euclidean correlator in Eq.~\eqref{eq:spectral_repr} is functionally identical to the Laplace transform. Of particular interest is the frequency shifting rule:
\begin{equation}
  G'(\tau) = e^{\Delta\cdot\tau}G(\tau)  
  \overset{\mathcal{L}}{\Longrightarrow}
  \rho'(\omega) = \rho(\omega+\Delta),
    \label{eq:laplace shift rule}
\end{equation}
where $\Delta$ is the shift parameter and $\rho(\omega)$
is the Laplace transform of $G(\tau)$.
However, the Backus-Gilbert spectral function, $\hat{\rho}(\omega_0)$,
will only be the true Laplace transform of $G(\tau)$ in the
limit that $A(\omega,\omega_0) = \delta(\omega-\omega_0)$,
see Eq.~\eqref{eq:bg}, which is not achievable on a finite system.
We also note that the
averaging functions $A(\omega,\omega_0)$ are invariant under the
shift\footnote{Ignoring information introduced during regularisation,
see Eq.~\eqref{eq:white}.}.
This means that because the shifted spectrum's ground state
features are closer to $\omega_\text{min}$, it will have
better resolution.
In this way, the Laplace shift should play a similar role to
the $\omega_\text{min}$ shift.
This is tested and confirmed in Fig.~\ref{fig:laplace_wmin_comparison}
where the Laplace
shifted spectrum's mass and FWHM is shown by the dashed curve.
This nicely overlays the results obtained from the
$\omega_\text{min}$ shift.

\begin{figure}[t]
    \centering
    \includegraphics[width=0.495\linewidth]{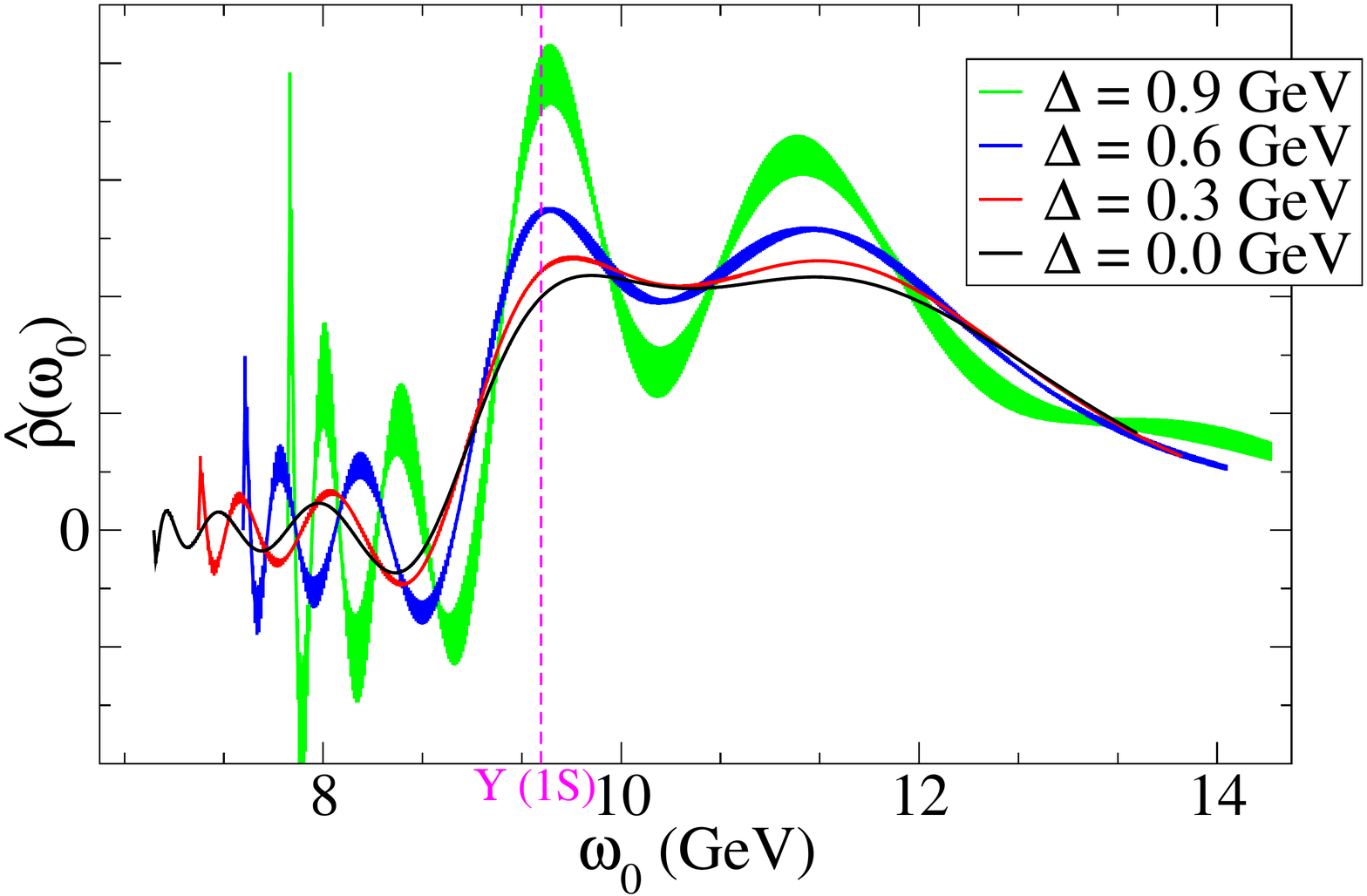}
    \includegraphics[width=0.495\linewidth]{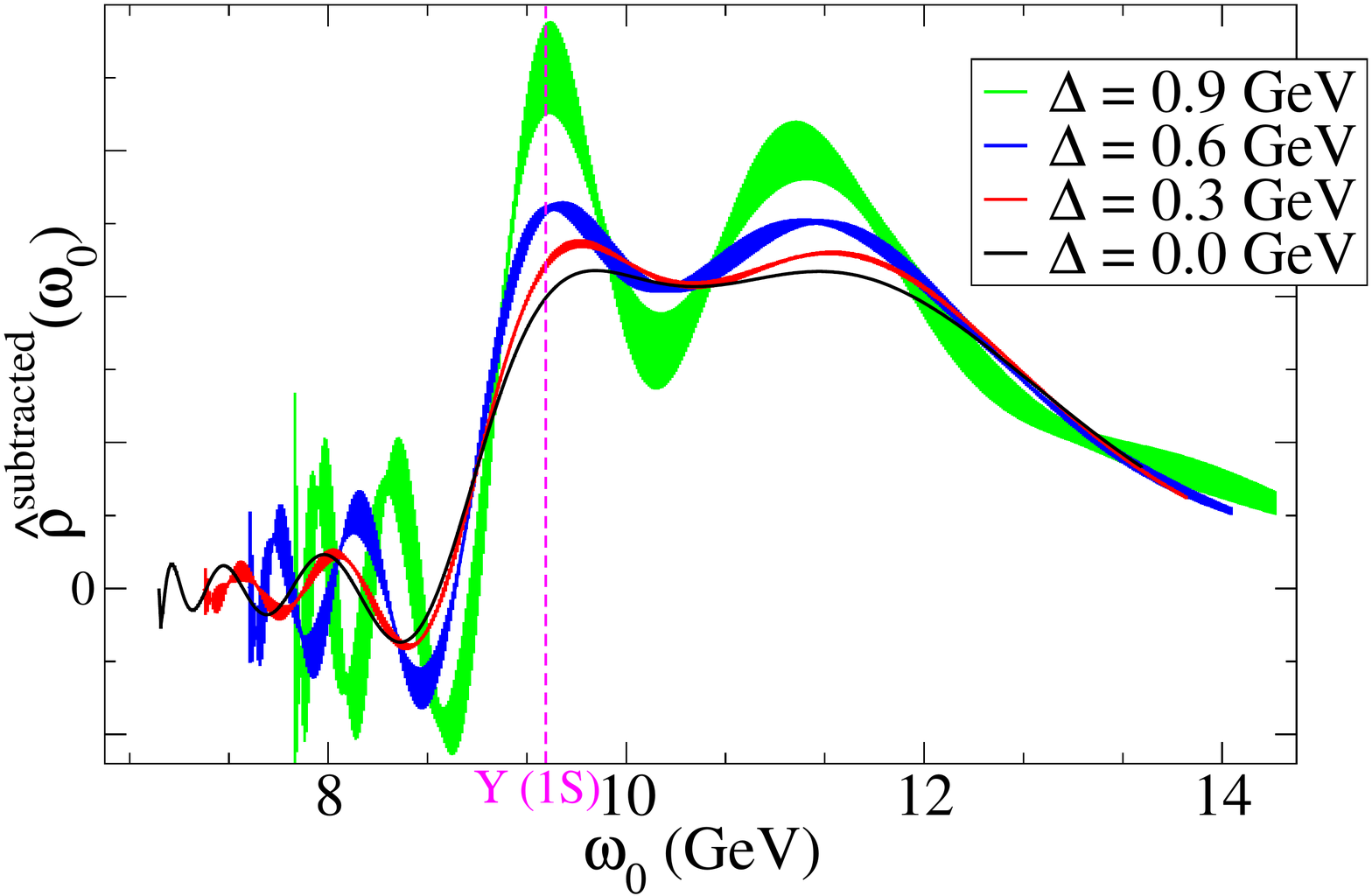}
    \caption{Left: Backus-Gilbert spectral functions at $T= 187$ MeV for a variety of Laplace shifts, $\Delta$, showing enhanced resolution as $\Delta$ increases. The magenta line is the experimental value of the $\Upsilon$ mass \cite{Zyla2020}. Right: The same spectral functions, but this time with noise subtraction using Eq.~\eqref{eq:rho-subtracted}. This shows a reduction in the unphysical features in the energy range below the ground state.}
    \label{fig:laplace shift}
\end{figure}

In Fig.\ \ref{fig:laplace shift} (left) the Laplace-shifted spectral 
functions are plotted for a variety of $\Delta$.
The enhanced resolution of the ground state feature as $\Delta$
increases can clearly be seen.
However, this comes at the expense of larger unphysical features
in the small-$\omega$ region.
This behaviour can be suppressed via an empirical noise subtraction 
procedure. The expression
\begin{equation}
    \bar{A}(\omega_0) = \int_{\omega_\text{min}}^{\omega_\text{max}} A(\omega,\omega_0)~d\omega
    \label{eq:laplace noise fit}
\end{equation}
is the Backus-Gilbert prediction for a flat spectrum,
$\rho(\omega) = \text{constant}$.
It is observed to closely match the unphysical small-$\omega$ 
fluctuations.
We use this to define the noise-subtracted spectral function,
\begin{equation}
    \hat{\rho}^\text{subtracted}(\omega_0) =
    \hat{\rho}(\omega_0) - \gamma \bar{A}(\omega_0),
    \label{eq:rho-subtracted}
\end{equation}
where $\gamma$ is tuned to minimise the small-$\omega$ fluctuations.
Fig.~\ref{fig:laplace shift} (right) plots $\hat{\rho}^\text{subtracted}$ which shows a reduction in the unphysical features below the ground state confirming
the usefulness of this noise subtraction procedure.

{\bf Assessment of the resolving power} --
\label{sec:resolving power}
%
%
From Eq.~\eqref{eq:bg}, the resolution function,
$\hat{\rho}(\omega_0) = A(M,\omega_0)$, 
is the Backus-Gilbert prediction for the case of a
spectral function $\rho(\omega) = \delta(\omega-M)$.
It therefore gives us a measure of the narrowest feature
that the method can resolve.

We determine the FWHM of $A(M,\omega_0)$,
using the same fit and rejection criteria as outlined 
in Sec.\ \ref{sec:overview}.
Fig.~\ref{fig:mass_width_comparison} (Right) depicts
these resolution FWHM values for the range of $T$ considered
for both the local and smeared cases.
These are a band of values because there is an ensemble of
$\omega_\text{min}$ and $\alpha$ values included,
see Sec.\ \ref{sec:overview}.
Since the Backus-Gilbert results do not sit above these
resolution band, we conclude that the method is unable
to resolve the width of the ground state.

\begin{figure}
    \centering
    \includegraphics[width=0.5\linewidth]{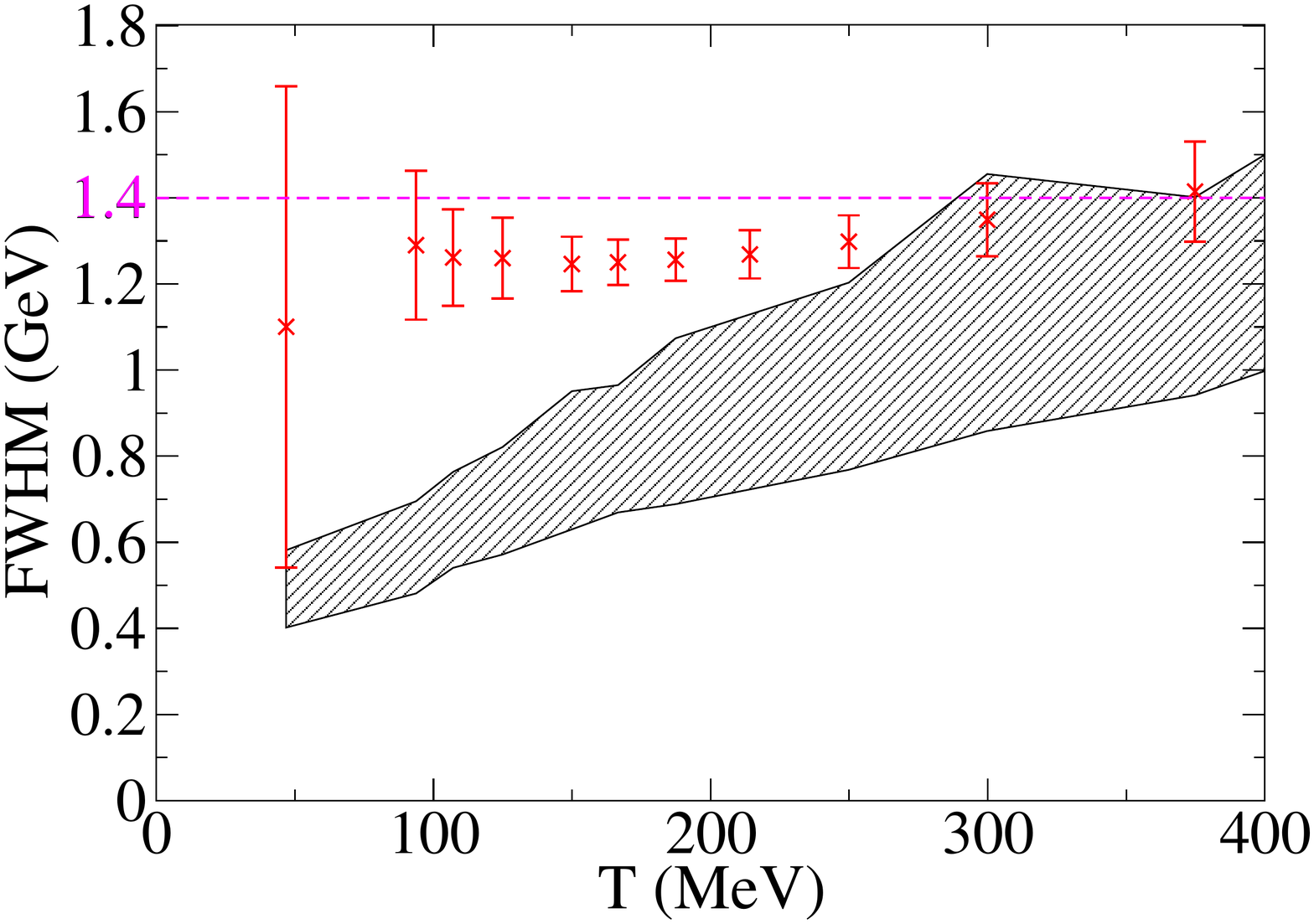}
    \caption{Plot of the FWHM obtained from the Backus-Gilbert method applied to a test spectrum of width $\sigma \simeq 1.4$ GeV. The same values of $N_\tau$ were used as in Table \ref{tab:Ntau_correlators}. Note that the large error bar for the smallest temperature is due to a high number of rejected fits (see Sec.\ \ref{sec:overview}).
    The grey region represents the resolution band.
    }
    \label{fig:gauss_width_test}
\end{figure}

We explore these issues further by applying the Backus-Gilbert
method to a test spectrum
consisting of a single, broad Gaussian of width
$\sigma \simeq1.4$ GeV centred at $M_\Upsilon \sim 9.4$ GeV.
The same $N_\tau$ range as in
Table \ref{tab:Ntau_correlators} was used, with
the same coefficient set, 
$c_\tau(\omega_0)$ as the local $\Upsilon$ data.
The results of the Backus-Gilbert FWHM are shown in 
Fig.~\ref{fig:gauss_width_test} together with the resolution band.
As can be seen, the width estimate exceeds the resolution for
$T \lesssim 300$ MeV.
This behaviour is what one would ought to expect if
the method was indeed capable of resolving the feature width
and contrasts with the results we find in 
Fig.~\ref{fig:mass_width_comparison}.
It is for this reason that we believe our estimates of the ground
state width instead represent an upper bound only.

\section{Summary}

We have presented preliminary results for the ground state mass and  width for the $\Upsilon$ meson, using both local and smeared quark sources. Since the widths are found to be consistent with the minimum resolvable width from the method, we consider our estimates of the ground state width to be upper bounds only. 

 We have also shown how the Laplace frequency shift transform or a shift of the energy window may be used to improve the resolving power of the method, with the caveat that a noise reduction routine must be employed to control unphysical oscillations in the small energy region in the former case.
 
\acknowledgments 

This work is supported by STFC grant ST/T000813/1.
SK is supported by the National Research Foundation of Korea under grant NRF-2021R1A2C1092701 funded by the Korean government (MEST).
BP has been supported by a Swansea University Research Excellence Scholarship (SURES).
This work used the DiRAC Extreme Scaling service at the University of Edinburgh, operated by the Edinburgh Parallel Computing Centre on behalf of the STFC DiRAC HPC Facility (www.dirac.ac.uk). This equipment was funded by BEIS capital funding via STFC capital grant ST/R00238X/1 and STFC DiRAC Operations grant ST/R001006/1. DiRAC is part of the National e-Infrastructure.
This work was performed using PRACE resources at Cineca via grants 2015133079 and 2018194714.
We acknowledge the support of the Supercomputing Wales project, which is part-funded by the European Regional Development Fund (ERDF) via Welsh Government,
and the University of Southern Denmark for use of computing facilities.
We are grateful to the Hadron Spectrum Collaboration for the use of their zero temperature ensemble.

\bibliographystyle{JHEP}
\bibliography{bg_refs}

\end{document}